\documentclass[prb,aps,twocolumn,showkeys,amsmath,amsfonts,floatfix,superscriptaddress,nofootinbib]{revtex4}
\usepackage{amssymb,amsmath,amstext}                
\usepackage{graphicx}                                               
\usepackage{tikz-network}
\usepackage{epstopdf}                                               
\usepackage{color}
\usepackage{bm}
\usepackage{appendix}                                              
\usepackage[latin2]{inputenc}
\usepackage{ulem}
\usepackage{latexsym}
\usepackage{tabularx}
\usepackage{afterpage}
\usepackage{nonfloat}
\usepackage[colorlinks=true,citecolor=blue,linkcolor=magenta]{hyperref}
\usepackage{lipsum}

\def\be{\begin{equation}}
\def\ee{\end{equation}}
\def\bea{\begin{eqnarray}}
\def\eea{\end{eqnarray}}
\def\bi{\begin{itemize}}
\def\ei{\end{itemize}}

\newcommand{\bra}[1]{\mbox{$\langle #1 |$}}
\newcommand{\ket}[1]{\mbox{$| #1 \rangle$}}
\newcommand{\braket}[2]{\mbox{$\langle #1  | #2 \rangle$}}

\begin{document}

\title{ Truncating loopy tensor networks by zero-mode gauge fixing:\\
         the $Z_2$ lattice gauge theory at finite temperature }

\newcommand{\affilju}{
             Jagiellonian University,
             Faculty of Physics, Astronomy and Applied Computer Science,
             Institute of Theoretical Physics,
             ul. \L{}ojasiewicza 11, 30-348 Krak\'ow, Poland
             }

\newcommand{\affildoc}{
             Jagiellonian University,
             Doctoral School of Exact and Natural Sciences,
             ul. \L{}ojasiewicza 11, 30-348 Krak\'ow, Poland
             }

\newcommand{\affilkac}{
             Jagiellonian University,
             Mark Kac Center for Complex Systems Research,
             ul. \L{}ojasiewicza 11, 30-348 Krak\'ow, Poland
             }

\author{Jacek Dziarmaga}\affiliation{\affilju}\affiliation{\affilkac}

\date{\today}

\begin{abstract}
    Loopy tensor networks exhibit internal correlations that often render their compression inefficient. We show that even local bond optimization can more effectively exploit locally available information about relevant loop correlations. By cutting a bond, we define a set of states whose linear dependence can be identified through a zero mode of the states' metric tensor and used to truncate the bond dimension.
    In the absence of an exact zero mode, a linear combination of a small number of the lowest modes can instead be optimized to provide the optimal approximation to a zero mode.
    The truncation does not require prior gauge fixing.
    The method is applied to the two-dimensional finite-temperature $Z_2$ lattice gauge theory, whose thermal-state purification is represented by an infinite projected entangled-pair state (iPEPS).
\end{abstract}


\maketitle


\section{Introduction}
\label{sec:intro}

Understanding strongly correlated quantum many-body systems remains a long-standing challenge, especially in two spatial dimensions (2D), where exact diagonalization is restricted to small system sizes and quantum Monte Carlo methods are hindered by the notorious sign problem. This difficulty can be alleviated by tensor network (TN) methods, which provide efficient representations of typical ground states of quantum many-body systems~\cite{Verstraete_review_08,Orus_review_14,Nishino_review_2022}. These include matrix product states (MPS) in one dimension (1D)~\cite{fannes1992,schollwock_review_2011}, projected entangled pair states (PEPS) in 2D~\cite{nishino01,gendiar03,verstraete2004} and 3D~\cite{Vlaar2021,3D_Charkiv}, and the multiscale entanglement renormalization ansatz (MERA)~\cite{Vidal_MERA_07,Vidal_MERA_08,Evenbly_branchMERA_14,Evenbly_branchMERAarea_14}.
Although MPS are highly powerful in 1D, owing to their canonical structure, their application in 2D is limited to relatively small system sizes. This restriction does not apply to PEPS~\cite{nishino01,gendiar03,verstraete2004, Murg_finitePEPS_07,Cirac_iPEPS_08,Xiang_SU_08,Orus_CTM_09,fu,Lubasch_conditioning,Corboz_varopt_16, Vanderstraeten_varopt_16, Fishman_FPCTM_17, Xie_PEPScontr_17, Corboz_Eextrap_16, Corboz_FCLS_18, Rader_FCLS_18, Rams_xiD_18, Hasik}, which constitute their natural generalization to higher dimensions. However, in the absence of an efficiently tractable canonical form~\cite{canonical_PEPS}, their expressive power may remain underutilized. The presence of closed loops in PEPS renders local tensor optimization less effective, as such procedures do not fully capture correlations circulating along these loops. In this work, we demonstrate that even local optimization can be improved to better exploit the locally accessible information about relevant loop correlations.

\begin{figure}[t!]
\includegraphics[width=0.9999\columnwidth]{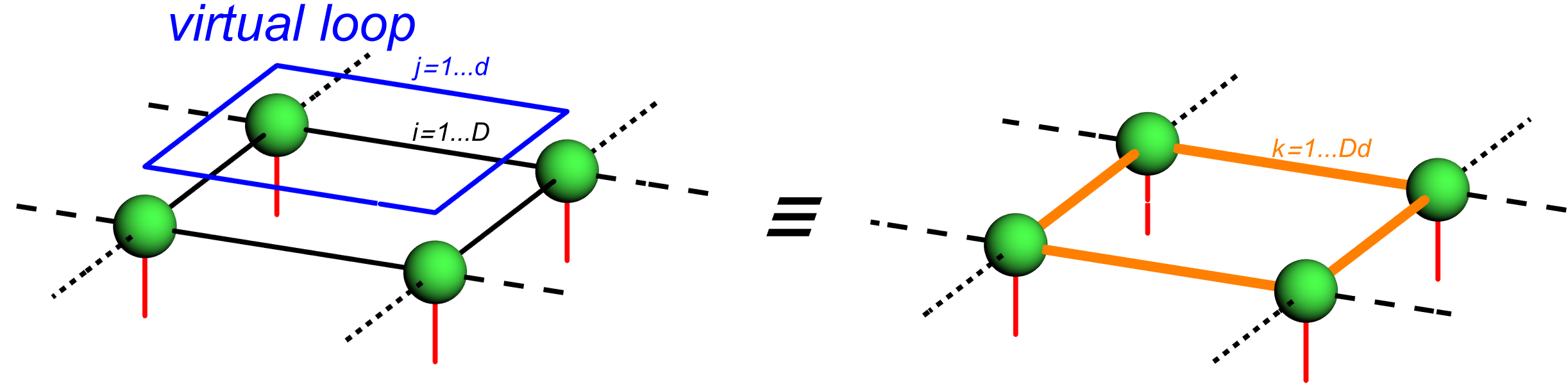}
\caption{{\bf Virtual entanglement loop. ---}
On the left, a four-tensor plaquette embedded in a larger tensor network (TN) via the dashed indices. The tensors are contracted along bond indices (black) of bond dimension $D$. In addition, an index loop carries a virtual index $j$ (blue) that is decoupled from any physical index (red). The TN state can be written as a sum over $j=1\ldots d$, $\ket{\rm TN}=\sum_{j=1}^d \ket{\psi_j}$, where all quantum states $\ket{\psi_j}$ are identical and proportional to the TN state.
On the right, the same plaquette after merging the indices $i$ and $j$ into a single index $k=1\ldots Dd$ (orange). The resulting bond dimension is larger by a factor of $d$ than required to represent the TN state. Any single component state $\ket{\psi_j}$, with the smaller bond dimension $D$, suffices to represent the same state, i.e., $\ket{\rm TN}\propto \ket{\psi_j}$.
}\label{fig:virtual}
\end{figure}

The nature of this issue can be illustrated by the example in Fig.~\ref{fig:virtual}. A virtual entanglement loop encircles a plaquette and is decoupled from the physical indices; nevertheless, it effectively inflates the bond dimensions along the plaquette edges. While this idealized example could, in principle, be remedied in several ways, virtual loops encountered in practice are more elusive: they are neither fully decoupled nor sharply defined. Accordingly, rather than pursuing a precise definition, we adopt a pragmatic approach. We select a bond in the tensor network that requires truncation, open it for inspection, and determine an optimal truncation of its bond dimension.
This paper is a follow-up to Ref.~\onlinecite{loops1}, in which this approach was introduced and its advantage was demonstrated by a series of examples. 
While the overall philosophy remains the same, the present formulation significantly simplifies the algorithm and renders it effectively gauge invariant under local transformations of the gauge indices.

The paper is organized as follows. In Sec.~\ref{sec:gauge}, we identify a gauge freedom arising from the linear dependence of the states constituting the tensor network. In Sec.~\ref{sec:zmt}, this freedom is exploited to truncate the bond dimension. In Sec.~\ref{sec:beyond}, the zero-mode truncation (ZMT) is generalized beyond exact zero modes. In Sec.~\ref{sec:invariance}, the invariance of the proposed algorithm under standard tensor-network gauge transformations is discussed. In Sec.~\ref{sec:Z2}, the method is applied to thermal states of the $Z_2$ lattice gauge theory on the square lattice. Section~\ref{sec:concl} concludes the paper.

\begin{figure}[t!]
\includegraphics[width=0.9999\columnwidth]{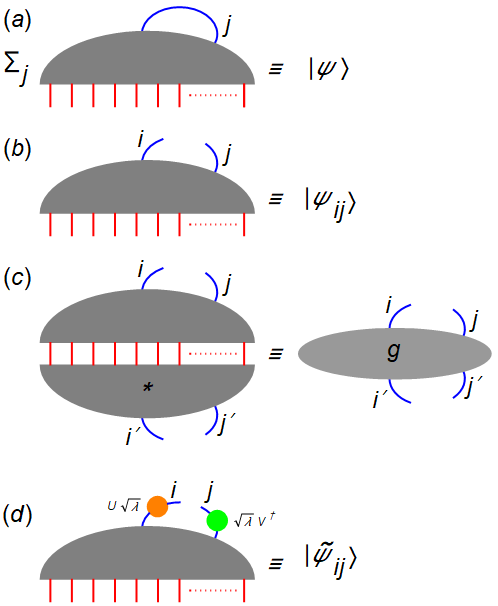}
\caption{{\bf Bond zero modes. ---}
In (a), the gray semi-ellipse denotes a tensor network (TN) state $\ket{\psi}$. The red lines correspond to physical indices. All internal bond indices are implicit, except for an explicit summation over the index $j$, indicated by the blue line.
In (b), this bond index line is cut to define the states $\ket{\psi_{ij}}$.
In (c), the overlaps between these states and their conjugates define the metric tensor in \eqref{eq:g_ij}.
In (d), the singular value decomposition \eqref{eq:UlambdaV} is absorbed into the TN, yielding the new states defined in \eqref{eq:tilde_psi}.
}\label{fig:psi}
\end{figure}

\section{Zero mode gauge freedom}
\label{sec:gauge}

Figure \ref{fig:psi}(a) shows a tensor network with an explicit summation over one of its bond indices. The TN can be written as
\be
\ket{\psi}=\sum_{i,j=1}^D \delta_{ij}\ket{\psi_{ij}},
\label{eq:psi_ij}
\ee
where the states $\ket{\psi_{ij}}$ are defined in Fig. \ref{fig:psi}(b). Their overlaps, shown in Fig. \ref{fig:psi}(c), define a metric tensor
\be
g_{ij,i'j'}=\braket{\psi_{ij}}{\psi_{i'j'}}.
\label{eq:g_ij}
\ee
Suppose that the metric possesses a zero mode $Z$ satisfying
\be 
\sum_{i',j'=1}^D g_{ij,i'j'} Z_{i'j'} = 0.
\ee
The existence of such a zero mode implies a linear dependence among the states $\ket{\psi_{ij}}$,
\be
\sum_{i,j=1}^D Z_{ij} \ket{\psi_{ij}} = 0,
\ee
which, in turn, provides a gauge freedom that allows one to rewrite \eqref{eq:psi_ij} as
\be
\ket{\psi}=
\sum_{i,j=1}^D
\left(
\delta_{ij} + z Z_{ij} \right)
\ket{\psi_{ij}},
\label{eq:psi_z_gen}
\ee
with an arbitrary parameter $z$. This freedom, arising from the linear dependence, can be exploited to truncate the bond dimension.

\section{Zero mode truncation}
\label{sec:zmt}

For instance, we can fix $z=-E_{\rm max}^{-1}$ in \eqref{eq:psi_z_gen}, where $E_{\rm max}$ denotes the eigenvalue of $Z_{ij}$ with the largest magnitude, such that $\delta_{ij} + z Z_{ij}$ becomes singular. Upon truncating $\lambda_D=0$ in its singular value decomposition,
\be
\delta_{ij} - E_{\rm max}^{-1} Z_{ij} = \sum_{k=1}^{D-1} U_{ik} \lambda_k V^*_{jk},
\label{eq:UlambdaV}
\ee
we can rewrite \eqref{eq:psi_z_gen} as
\be
\ket{\psi} = \sum_{k=1}^{D-1} \lambda_k \sum_{i,j=1}^D U_{ik} V^*_{jk} \ket{\psi_{ij}}.
\ee
Defining new states
\be
\ket{\tilde\psi_{ij}} =
\sqrt{\lambda_i \lambda_j}
\sum_{i',j'=1}^D U_{i'i} V^*_{j'j} \ket{\psi_{i'j'}},
\label{eq:tilde_psi}
\ee
as shown in Fig. \ref{fig:psi}(d), the TN state in \eqref{eq:psi_ij} becomes
\be
\ket{\psi}=
\sum_{i,j=1}^{D-1}
\delta_{ij}
\ket{\tilde\psi_{ij}}.
\label{eq:tilde_psi_ij}
\ee
Compared to the original expression \eqref{eq:psi_ij}, the bond dimension is reduced from $D$ to $D-1$. If the new metric tensor
$
\tilde{g}_{ij,i'j'}=\braket{\tilde\psi_{ij}}{\tilde\psi_{i'j'}}
$
again possesses a zero mode, the truncation procedure can be iterated, further reducing the bond dimension from $D-1$ to $D-2$.

\section{Beyond zero modes}
\label{sec:beyond}

When the smallest eigenvalue of the metric tensor is nonzero but small, the corresponding eigenmode $Z_{ij}$ can still be used for truncation in the manner as an exact zero mode. The truncation error is given by the squared norm of the difference between the tensor-network states before and after truncation; cf. \eqref{eq:psi_ij} and \eqref{eq:psi_z_gen} with $z=-1/E_{\rm max}$. It takes the form
\be
f = \frac{N}{|E_{\rm max}|^2}.
\label{eq:f}
\ee
Here, $N$ denotes the quadratic form defined by the metric tensor,
\be
N = \sum_{i,j,i',j'=1}^D Z^*_{ij}~ g_{ij,i'j'}~ Z_{i'j'},
\label{eq:N}
\ee
while $E_{\rm max}$ is the eigenvalue of the matrix $Z_{ij}$ with the largest magnitude. The denominator in \eqref{eq:f} motivates this choice of eigenvalue.

As Eq.~\eqref{eq:f} does not, in fact, require $Z_{ij}$ to be an exact eigenmode, it is advantageous to consider a more general ansatz in which $Z_{ij}$ is expressed as a linear combination of a small number of the lowest-lying eigenmodes, thereby increasing $|E_{\rm max}|$:
\be
Z_{ij} = \sum_{m=1}^{\kappa} \alpha_m Z^{m}_{ij}.
\label{eq:subspace}
\ee
Here, $Z^{m}_{ij}$ denotes the $m$-th lowest eigenmode, and $\alpha_m$ are variational coefficients optimized by minimizing the cost function \eqref{eq:f}.

To carry out this optimization, we require the gradient of the cost function \eqref{eq:f} with respect to a general $Z_{ij}$:
\bea 
G_{ij} \equiv 
\frac{\partial f}{\partial Z^*_{ij}} = 
\frac{1}{|E_{\rm max}|^2} 
\left( 
\frac{\partial N}{\partial Z^*_{ij}} -\frac{N}{E_{\rm max}^*} \frac{\partial E_{\rm max}^*}{\partial Z^*_{ij}} 
\right).
\eea 
Using the eigendecomposition
\be
Z_{ij} = R_i E_{\rm max} L_j + \dots,
\ee
and hence
\be
\frac{\partial E_{\rm max}^*}{\partial Z_{ij}} = L_i^* R_j^*. 
\ee
Here $\sum_{j=1}^D L_j R_j =1$. Combining this with the derivative of \eqref{eq:N}, we arrive at the gradient
\bea
G_{ij} =
\frac{1}{|E_{\rm max}|^2}
\sum_{i',j'=1}^D
\left(
g_{ij,i'j'} - f~p_{ij,i'j'}
\right)
Z_{i'j'},
\label{eq:G}
\eea
where $p_{ij,i'j'} = \left(L^*_i R^*_j\right)\left(L_{i'} R_{j'}\right)$ is proportional to a Hermitian projector. Notably, Eq.~\eqref{eq:G} holds for any non-degenerate eigenvalue $E_{\rm max}$, not necessarily the leading one.
Within the $\kappa$-dimensional subspace defined in \eqref{eq:subspace}, the gradient reduces to
\be
G_m \equiv
\frac{\partial f}{\partial \alpha_m^*} =
\sum_{i,j=1}^D
\frac{\partial Z^*_{ij}}{\partial \alpha^*_m}
\frac{\partial f}{\partial Z^*_{ij}} =
\sum_{i,j=1}^D
Z^{*m}_{ij} G_{ij},
\label{eq:Gm}
\ee
which can be used to optimize the amplitudes $\alpha_m$.

These considerations lead to the following algorithm:
\bi
\item[1)] Find the $\kappa$ lowest eigenmodes of the metric tensor $g$. To make it non-singular, add a            small positive constant: $g\to g+\delta$;
\item[2)] Select the eigenmode $Z^{m_{\rm min}}_{ij}$ with the smallest truncation error \eqref{eq:f}   and initialize $\alpha_{m_{\rm min}}=1$ and $0$ otherwise;
\item[3)] Optimize the amplitudes $\alpha_m$ using a conjugate gradient method with the gradient   \eqref{eq:Gm};
\item[4)] Use the optimal $Z_{ij}$ to perform the truncation as in Sec.~\ref{sec:zmt}.
\ei

For a real tensor network, a real matrix $Z_{ij}$ does not, in general, have purely real eigenvalues; rather, complex eigenvalues may occur in conjugate pairs. To ensure that the subtraction in \eqref{eq:UlambdaV} remains real, $E_{\rm max}$ is taken to be the real eigenvalue of largest magnitude. To guarantee the existence of at least one real eigenvalue, $\kappa$ can be chosen to be odd. The remainder of the algorithm then proceeds unchanged.

For a symmetric tensor network, $Z_{ij}$ is recast as $Z^s_{i_s j_s}$, where $s$ labels the symmetry sector and $i_s, j_s$ enumerate the states within sector $s$. Accordingly, the metric tensor $g_{ij,i'j'}$ takes the form
$
g^{s,s'}_{i_s j_s, i'_{s'} j'_{s'}}.
$
The optimal $Z^s_{i_s j_s}$ is then diagonalized independently in each sector $s$; consequently, the maximal eigenvalue $E_{\rm max}$ resides in a well-defined sector. The truncation procedure reduces the dimension of this sector by one.

\section{Gauge invariance}
\label{sec:invariance}

For the optimized $Z_{ij}$ in \eqref{eq:subspace}, the matrix \eqref{eq:UlambdaV} - which is inserted into a bond of a tensor network - can be diagonalized as
\be
\delta_{ij} - E_{\rm max}^{-1}~Z_{ij} = \sum_{k=1}^{D-1} S^{-1}_{ik} \mu_k S_{kj}.
\label{eq:UlambdaVdiag}
\ee
The matrix $S$ can be interpreted as a standard tensor-network gauge transformation: inserting the identity $S^{-1}S$ on a bond leaves the TN unchanged. In this gauge, the truncation is implemented by inserting a diagonal matrix containing $D-1$ nonzero eigenvalues $\mu$. 

The proposed algorithm, consisting of identifying the optimal subspace followed by conjugate-gradient optimization, is therefore equivalent to a simultaneous optimization of the gauge transformation $S$ and the truncated spectrum $\mu$. Its advantage lies in the separation of its two steps: the first is a non-variational identification of a small number $\kappa$ of relevant variational parameters, while the second is a variational optimization over this reduced parameter set. The reduction is meant to minimize the probability of getting trapped in a local minimum.

Suppose now that a standard gauge transformation ${\cal G}^{-1}{\cal G}$ is inserted on a bond prior to applying the algorithm. If we neglect the possibility of trapping in a local minimum for a given ${\cal G}$, the algorithm identifies the globally optimal insertion $\delta_{ij}-Z^{{\cal G}}_{ij}/E^{{\cal G}}_{\rm max}=S_{\cal G}^{-1}\mu_{\cal G}S_{\cal G}$. This global optimum is gauge equivalent to that obtained in the original gauge:
\be
{\cal G}^{-1} \left( 1 - Z^{{\cal G}}/E^{{\cal G}}_{\rm max} \right) {\cal G} = 1 - Z/E_{\rm max}.
\ee
It follows that the optimal eigenvalues are gauge invariant, $\mu_{\cal G}=\mu$, whereas the corresponding similarity transformation differs only by a gauge transformation, $S_{\cal G}=S{\cal G}^{-1}$. Owing to this invariance, no explicit gauge fixing is required prior to truncation.

\begin{figure}[t!]
\includegraphics[width=0.9999\columnwidth]{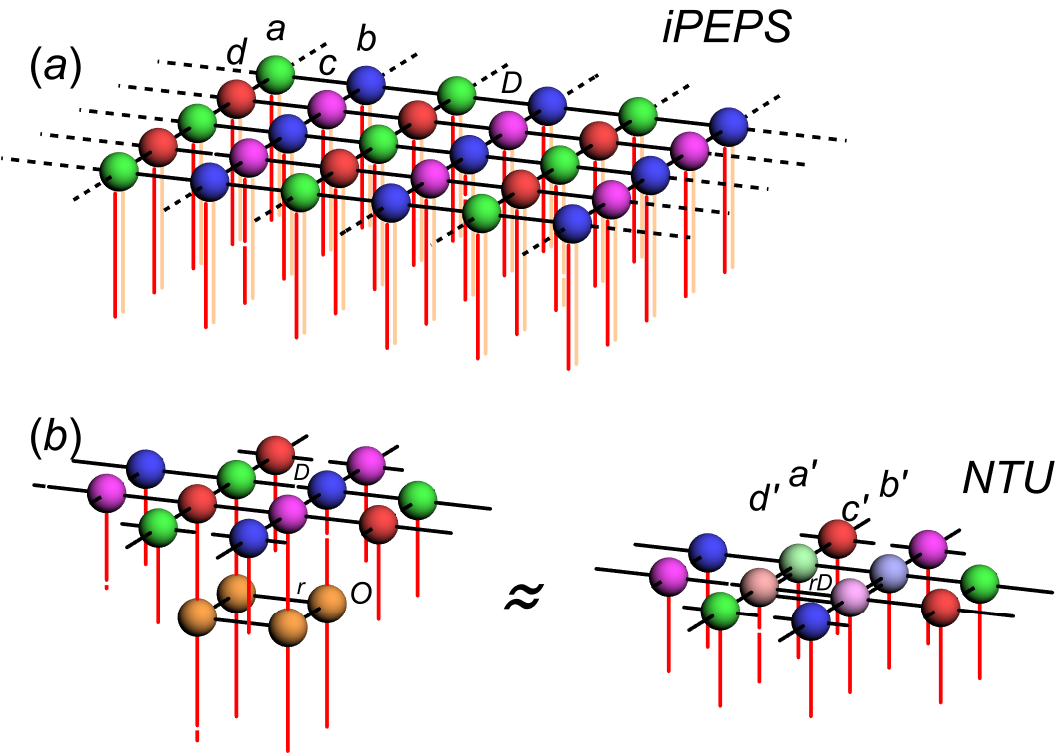}
\caption{
{\bf NTU for gauge field. }
In (a), we depict the infinite PEPS (iPEPS) tensor-network ansatz, consisting of four sublattice tensors, $a,b,c,d$. All bond indices have dimension $D$, while the red (orange) lines represent spin (ancilla) indices.
In (b), left, the plaquette evolution operator \eqref{eq:U_p}, expressed as a matrix-product operator (MPO) \eqref{eq:MPO}, is applied to the $abcd$ plaquette of iPEPS tensors. The MPO bond dimension is $r=2$. For clarity, the ancilla indices are omitted.
In (b), right, the bond $a'-b'$ is first truncated to dimension $D$ using zero-mode truncation. Subsequently, the tensors $a'$ and $b'$ are variationally optimized to minimize the difference between the target network (left) and the truncated network (right).
Upon convergence, the procedure is repeated for the bond $b'-c'$ within the corresponding local tensor environment. The same sequence of truncation and optimization is then applied to the bonds $c'-d'$ and $d'-a'$.
The truncation error $\delta$ is defined as the Frobenius norm of the difference between the target (left) and variational (right) tensor networks. In Fig.~\ref{fig:ntu_results} (a), we show $\delta$ after the initial zero-mode truncation (ZMT initial) and after the subsequent optimization (ZMT final). For comparison, the error obtained from an initial SVD-based truncation is also shown. All errors are averaged over the bonds $a'-b'$, $b'-c'$, $c'-d'$, and $d'-a'$ of the $abcd$ plaquette and the bonds $c'-d'$, $d'-b'$, $b'-a'$, and $a'-c'$ of the $cdba$ plaquette. 
}\label{fig:ntu_gauge}
\end{figure}

\begin{figure}[t!]
\includegraphics[width=0.94\columnwidth]{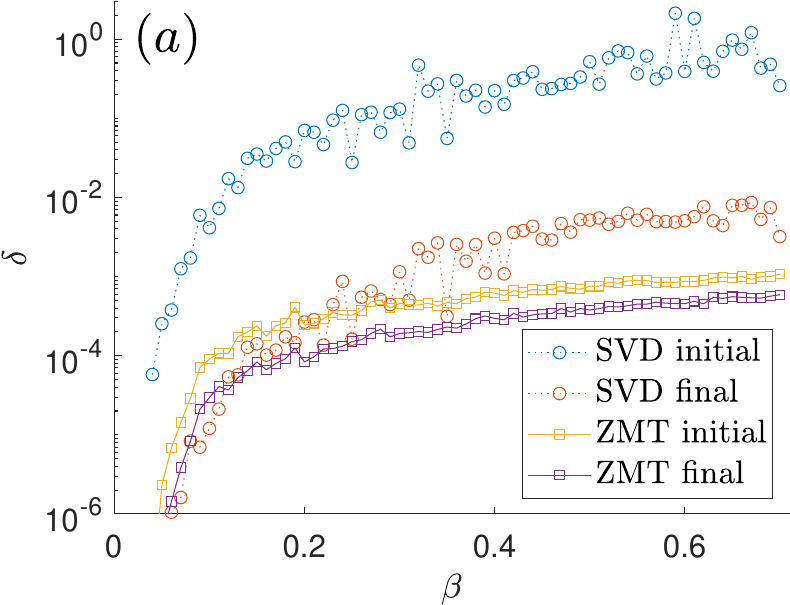}
\includegraphics[width=0.9\columnwidth]{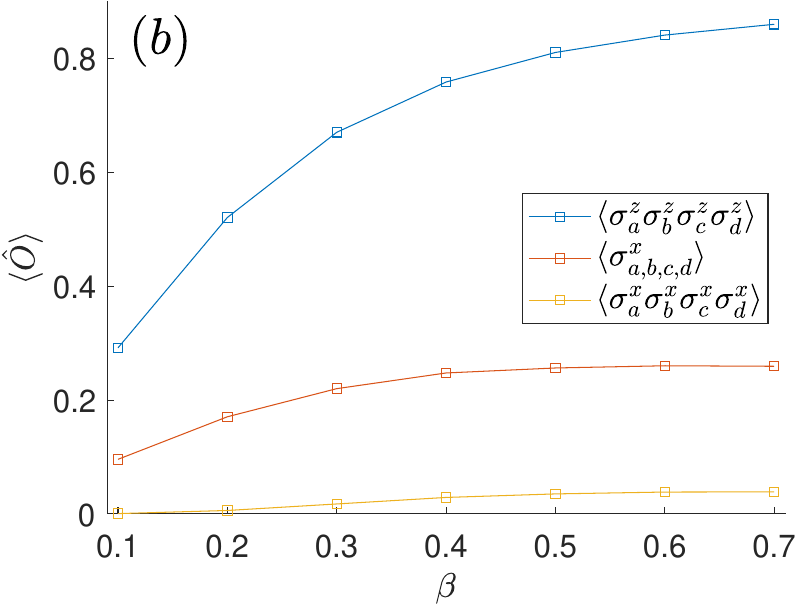}
\caption{
{\bf Gauge field - errors and observables. }
In (a), we present the average truncation error $\delta$, as defined in Fig.~\ref{fig:ntu_gauge}(b), as a function of the inverse temperature $\beta$. The data are shown immediately after the zero-mode truncation (ZMT initial) and after the subsequent optimization (ZMT final).
For comparison, the same panel also includes results obtained with the simple SVD truncation, shown both immediately after truncation (SVD initial) and after the subsequent optimization (SVD final).
Panel (b) displays selected nontrivial observables as a function of $\beta$.
Here $D=10$, $\kappa=5$, $\chi=40$, $d\beta=0.01$, and $g=3.04438$. A real tensor-network algorithm is employed, with $E_{\rm max}$ taken as the real eigenvalue of largest magnitude.
}\label{fig:ntu_results}
\end{figure}

\section{Thermal $Z_2$ lattice gauge theory}
\label{sec:Z2}

The $Z_2$ gauge-field Hamiltonian on an infinite square lattice can be written as
\be
H = H_m + H_e \equiv
-g\sum_p B_p
-\sum_s \sigma^x_s.
\ee
In the magnetic term $H_m$, the index $p$ runs over the white $2\times 2$ plaquettes of a checkerboard tiling. The plaquette operator is defined as $B_p=\sigma^z_{p_1} \sigma^z_{p_2} \sigma^z_{p_3} \sigma^z_{p_4}$, where $p_i$ label the four corner sites of plaquette $p$. In the electric term $H_e$, the index $s$ runs over all lattice sites. We employ a dual lattice construction, in which fermionic degrees of freedom would reside on sites located at the centers of the black plaquettes. In the absence of charges, the Gauss-law constraint imposes $A_p\equiv\sigma^x_{p_1}\sigma^x_{p_2}\sigma^x_{p_3}\sigma^x_{p_4}=1$ on every black plaquette $p$.

To represent the thermal Gibbs state $\rho$, we associate each lattice spin ${\cal S}_s$ with an ancilla spin ${\cal A}_s$ and employ a purification,
\be
\rho(\beta)=
e^{-\beta H}=
{\rm Tr}_{\cal A} 
\ket{\psi_{\cal SA}(\beta)} \bra{\psi_{\cal SA}(\beta)}.
\ee
The purified state is obtained via imaginary-time evolution,
$\ket{\psi_{\cal SA}(\beta)} = e^{-\frac12\beta H}\ket{\psi_{\cal SA}(0)}$, starting from a product state over lattice sites.
\be
\ket{\psi_{\cal SA}(0)} \propto
\prod_s
\left(
\ket{\uparrow_{{\cal S}_s} }\ket{\uparrow_{{\cal A}_s} }
+
\ket{\downarrow_{{\cal S}_s} }\ket{\downarrow_{{\cal A}_s} }
\right).
\ee
The Gauss-law constraint is enforced by inserting a projector ${\cal P}$ into the expectation value of a charge-conserving spin operator ${\cal O}$,
\be
\bra{\psi_{\cal SA}(\beta)} {\cal O} {\cal P} \ket{\psi_{\cal SA}(\beta)}.
\ee
In the chargeless sector, the projector takes the form ${\cal P}\propto\prod_p (1+A_p)$, where $p$ runs over the black plaquettes.

As in many previous works~\cite{CzarnikDziarmagaCorboz,ntu,Hubbard_Sinha,Sinha_Wietek_Hubbard,tJ_Zhang},
the purification is represented by an iPEPS, as illustrated in Fig.~\ref{fig:ntu_gauge}(a), with four sublattice tensors $a\ldots d$ and bond dimension $D$. The projector ${\cal P}$ is expressed as an iPEPO with bond dimension $2$. The expectation value is evaluated using the corner transfer matrix renormalization group~\cite{gapCorboz}, by computing the overlap between the iPEPS $\ket{\psi_{\cal SA}(\beta)}$ with bond dimension $D$ and the iPEPS ${\cal P} \ket{\psi_{\cal SA}(\beta)}$ with bond dimension $2D$.

The evolution operator $e^{-\frac12\beta H}$ is decomposed into small imaginary-time steps $d\beta$, with each step approximated using a second-order Suzuki-Trotter decomposition,
\be
e^{-\frac12 d\beta H}
\approx
e^{-\frac14 d\beta H_e}
e^{-\frac12 d\beta H_m}
e^{-\frac14 d\beta H_e}.
\ee
The electric evolution operator factorizes into a product of local site operators,
\be
e^{-\frac14 d\beta H_e} = \prod_s e^{\frac14 d\beta \sigma^x_s}.
\ee
Accordingly, each iPEPS tensor is updated by the local transformation $e^{\frac14 d\beta \sigma^x_s}$.
The magnetic evolution operator $e^{-\frac12 d\beta H_m}$ is a product of mutually commuting plaquette operators,
\bea
& &
U_p(d\beta) = e^{\varepsilon B_p }= \label{eq:U_p}\\
& &
1_{p_1} 1_{p_2} 1_{p_3} 1_{p_4} ~
\cosh\varepsilon~ +
\sigma^z_{p_1} \sigma^z_{p_2} \sigma^z_{p_3} \sigma^z_{p_4} ~
\sinh\varepsilon,
\nonumber
\eea
applied to the white plaquettes $p$. Here $\varepsilon=g d\beta/2$.This decomposition consists of two contributions. In the first, each tensor on the plaquette is acted on by $\left(\cosh\varepsilon\right)^{1/4} 1_{p_i}\equiv O_{p_i}^{1}(d\beta)$, while in the second it is acted on by $\left(\sinh\varepsilon\right)^{1/4} \sigma^z_{p_i}\equiv O_{p_i}^{2}(d\beta)$. The magnetic evolution operator can thus be expressed as a periodic matrix product operator (pMPO),
\be
U_p(d\beta) =
\sum_{j=1,2}
O_{p_1}^{j}(d\beta) O_{p_2}^{j}(d\beta) O_{p_3}^{j}(d\beta) O_{p_4}^{j}(d\beta),
\label{eq:MPO}
\ee
where $j$ is a virtual loop index running around the plaquette (see Fig.~\ref{fig:ntu_gauge}(b)). In each time step, these commuting pMPOs are first applied to all $a$-$b$-$c$-$d$ plaquettes, as shown in Fig.~\ref{fig:ntu_gauge}(b), and subsequently to all $c$-$d$-$b$-$a$ plaquettes.

The results of the evolution are summarized in Fig. \ref{fig:ntu_results}. The top panel shows the average truncation error per bond at each step. It displays both the initial truncation error after the zero-mode truncation (ZMT) and the final truncation error obtained after further optimization of the truncated tensors. As a benchmark, the same panel also shows the corresponding errors when a simple SVD truncation is used in place of ZMT. Notably, ZMT yields smaller final errors than SVD; moreover, at later stages of the evolution, even the initial ZMT error is lower than the final SVD error. The improvement of ZMT over SVD amounts to approximately one order of magnitude.

\section{Conclusion}
\label{sec:concl}

We presented a more efficient and effectively gauge-invariant version of the method introduced in Ref.~\onlinecite{loops1}. The method identifies (near-)zero modes associated with a bond in a tensor network and uses them to truncate the bond index by eliminating the (approximate) linear dependence of the states constituting the network.
The advantage of the algorithm was demonstrated for thermal states of the lattice gauge theory, where it yields a truncation error an order of magnitude smaller than that obtained with the standard optimization initialized by SVD truncation.

The data used for the figures in this article are openly available from the RODBUK repository at https://doi.org/10.57903/UJ/TRYKVW. 

\acknowledgments

I am indebted to Marek Rams, Ihor Sokolov and Yintai Zhang for stimulating discussions.
This research was
%
%
funded by the National Science Centre (NCN), Poland, under projects 2024/55/B/ST3/00626 (JD). 
%
%
This research was also supported by a grant from the Priority Research Area DigiWorld under the Strategic Programme Excellence Initiative at Jagiellonian University (JD).

\bibliographystyle{apsrev4-2}
\bibliography{KZref.bib}

\begin{thebibliography}{37}%
\makeatletter
\providecommand \@ifxundefined [1]{%
 \@ifx{#1\undefined}
}%
\providecommand \@ifnum [1]{%
 \ifnum #1\expandafter \@firstoftwo
 \else \expandafter \@secondoftwo
 \fi
}%
\providecommand \@ifx [1]{%
 \ifx #1\expandafter \@firstoftwo
 \else \expandafter \@secondoftwo
 \fi
}%
\providecommand \natexlab [1]{#1}%
\providecommand \enquote  [1]{``#1''}%
\providecommand \bibnamefont  [1]{#1}%
\providecommand \bibfnamefont [1]{#1}%
\providecommand \citenamefont [1]{#1}%
\providecommand \href@noop [0]{\@secondoftwo}%
\providecommand \href [0]{\begingroup \@sanitize@url \@href}%
\providecommand \@href[1]{\@@startlink{#1}\@@href}%
\providecommand \@@href[1]{\endgroup#1\@@endlink}%
\providecommand \@sanitize@url [0]{\catcode `\\12\catcode `\$12\catcode `\&12\catcode `\#12\catcode `\^12\catcode `\_12\catcode `\%12\relax}%
\providecommand \@@startlink[1]{}%
\providecommand \@@endlink[0]{}%
\providecommand \url  [0]{\begingroup\@sanitize@url \@url }%
\providecommand \@url [1]{\endgroup\@href {#1}{\urlprefix }}%
\providecommand \urlprefix  [0]{URL }%
\providecommand \Eprint [0]{\href }%
\providecommand \doibase [0]{https://doi.org/}%
\providecommand \selectlanguage [0]{\@gobble}%
\providecommand \bibinfo  [0]{\@secondoftwo}%
\providecommand \bibfield  [0]{\@secondoftwo}%
\providecommand \translation [1]{[#1]}%
\providecommand \BibitemOpen [0]{}%
\providecommand \bibitemStop [0]{}%
\providecommand \bibitemNoStop [0]{.\EOS\space}%
\providecommand \EOS [0]{\spacefactor3000\relax}%
\providecommand \BibitemShut  [1]{\csname bibitem#1\endcsname}%
\let\auto@bib@innerbib\@empty
\bibitem [{\citenamefont {Verstraete}\ \emph {et~al.}(2008)\citenamefont {Verstraete}, \citenamefont {Murg},\ and\ \citenamefont {Cirac}}]{Verstraete_review_08}%
  \BibitemOpen
  \bibfield  {author} {\bibinfo {author} {\bibfnamefont {F.}~\bibnamefont {Verstraete}}, \bibinfo {author} {\bibfnamefont {V.}~\bibnamefont {Murg}},\ and\ \bibinfo {author} {\bibfnamefont {J.}~\bibnamefont {Cirac}},\ }\href {https://doi.org/10.1080/14789940801912366} {\bibfield  {journal} {\bibinfo  {journal} {Adv. Phys.}\ }\textbf {\bibinfo {volume} {57}},\ \bibinfo {pages} {143} (\bibinfo {year} {2008})}\BibitemShut {NoStop}%
\bibitem [{\citenamefont {Or\'us}(2014)}]{Orus_review_14}%
  \BibitemOpen
  \bibfield  {author} {\bibinfo {author} {\bibfnamefont {R.}~\bibnamefont {Or\'us}},\ }\href {http://www.sciencedirect.com/science/article/pii/S0003491614001596} {\bibfield  {journal} {\bibinfo  {journal} {Ann. Phys. (Amsterdam)}\ }\textbf {\bibinfo {volume} {349}},\ \bibinfo {pages} {117 } (\bibinfo {year} {2014})}\BibitemShut {NoStop}%
\bibitem [{\citenamefont {Okunishi}\ \emph {et~al.}(2022)\citenamefont {Okunishi}, \citenamefont {Nishino},\ and\ \citenamefont {Ueda}}]{Nishino_review_2022}%
  \BibitemOpen
  \bibfield  {author} {\bibinfo {author} {\bibfnamefont {K.}~\bibnamefont {Okunishi}}, \bibinfo {author} {\bibfnamefont {T.}~\bibnamefont {Nishino}},\ and\ \bibinfo {author} {\bibfnamefont {H.}~\bibnamefont {Ueda}},\ }\href {https://doi.org/10.7566/JPSJ.91.062001} {\bibfield  {journal} {\bibinfo  {journal} {Journal of the Physical Society of Japan}\ }\textbf {\bibinfo {volume} {91}},\ \bibinfo {pages} {062001} (\bibinfo {year} {2022})},\ \Eprint {https://arxiv.org/abs/https://doi.org/10.7566/JPSJ.91.062001} {https://doi.org/10.7566/JPSJ.91.062001} \BibitemShut {NoStop}%
\bibitem [{\citenamefont {Fannes}\ \emph {et~al.}(1992)\citenamefont {Fannes}, \citenamefont {Nachtergaele},\ and\ \citenamefont {Werner}}]{fannes1992}%
  \BibitemOpen
  \bibfield  {author} {\bibinfo {author} {\bibfnamefont {M.}~\bibnamefont {Fannes}}, \bibinfo {author} {\bibfnamefont {B.}~\bibnamefont {Nachtergaele}},\ and\ \bibinfo {author} {\bibfnamefont {R.}~\bibnamefont {Werner}},\ }\href {http://dx.doi.org/10.1007/BF02099178} {\bibfield  {journal} {\bibinfo  {journal} {Comm. in Math. Phys.}\ }\textbf {\bibinfo {volume} {144}},\ \bibinfo {pages} {443} (\bibinfo {year} {1992})}\BibitemShut {NoStop}%
\bibitem [{\citenamefont {Sch\"ollwock}(2011)}]{schollwock_review_2011}%
  \BibitemOpen
  \bibfield  {author} {\bibinfo {author} {\bibfnamefont {U.}~\bibnamefont {Sch\"ollwock}},\ }\href {https://doi.org/http://dx.doi.org/10.1016/j.aop.2010.09.012} {\bibfield  {journal} {\bibinfo  {journal} {Ann. Phys.}\ }\textbf {\bibinfo {volume} {326}},\ \bibinfo {pages} {96} (\bibinfo {year} {2011})}\BibitemShut {NoStop}%
\bibitem [{\citenamefont {Nishino}\ \emph {et~al.}(2001)\citenamefont {Nishino}, \citenamefont {Hieida}, \citenamefont {Okunishi}, \citenamefont {Maeshima}, \citenamefont {Akutsu},\ and\ \citenamefont {Gendiar}}]{nishino01}%
  \BibitemOpen
  \bibfield  {author} {\bibinfo {author} {\bibfnamefont {T.}~\bibnamefont {Nishino}}, \bibinfo {author} {\bibfnamefont {Y.}~\bibnamefont {Hieida}}, \bibinfo {author} {\bibfnamefont {K.}~\bibnamefont {Okunishi}}, \bibinfo {author} {\bibfnamefont {N.}~\bibnamefont {Maeshima}}, \bibinfo {author} {\bibfnamefont {Y.}~\bibnamefont {Akutsu}},\ and\ \bibinfo {author} {\bibfnamefont {A.}~\bibnamefont {Gendiar}},\ }\href {http://ptp.oxfordjournals.org/content/105/3/409} {\bibfield  {journal} {\bibinfo  {journal} {Prog. Theor. Phys.}\ }\textbf {\bibinfo {volume} {105}},\ \bibinfo {pages} {409} (\bibinfo {year} {2001})}\BibitemShut {NoStop}%
\bibitem [{\citenamefont {Gendiar}\ \emph {et~al.}(2003)\citenamefont {Gendiar}, \citenamefont {Maeshima},\ and\ \citenamefont {Nishino}}]{gendiar03}%
  \BibitemOpen
  \bibfield  {author} {\bibinfo {author} {\bibfnamefont {A.}~\bibnamefont {Gendiar}}, \bibinfo {author} {\bibfnamefont {N.}~\bibnamefont {Maeshima}},\ and\ \bibinfo {author} {\bibfnamefont {T.}~\bibnamefont {Nishino}},\ }\href {http://ptp.oxfordjournals.org/content/110/4/691} {\bibfield  {journal} {\bibinfo  {journal} {Prog. Theor. Phys.}\ }\textbf {\bibinfo {volume} {110}},\ \bibinfo {pages} {691} (\bibinfo {year} {2003})}\BibitemShut {NoStop}%
\bibitem [{\citenamefont {Verstraete}\ and\ \citenamefont {Cirac}(2004)}]{verstraete2004}%
  \BibitemOpen
  \bibfield  {author} {\bibinfo {author} {\bibfnamefont {F.}~\bibnamefont {Verstraete}}\ and\ \bibinfo {author} {\bibfnamefont {J.~I.}\ \bibnamefont {Cirac}},\ }\href {http://arxiv.org/abs/cond-mat/0407066} {\bibfield  {journal} {\bibinfo  {journal} {arXiv:cond-mat/0407066}\ } (\bibinfo {year} {2004})}\BibitemShut {NoStop}%
\bibitem [{\citenamefont {Vlaar}\ and\ \citenamefont {Corboz}(2021)}]{Vlaar2021}%
  \BibitemOpen
  \bibfield  {author} {\bibinfo {author} {\bibfnamefont {P.~C.~G.}\ \bibnamefont {Vlaar}}\ and\ \bibinfo {author} {\bibfnamefont {P.}~\bibnamefont {Corboz}},\ }\href {https://doi.org/10.1103/PhysRevB.103.205137} {\bibfield  {journal} {\bibinfo  {journal} {Phys. Rev. B}\ }\textbf {\bibinfo {volume} {103}},\ \bibinfo {pages} {205137} (\bibinfo {year} {2021})}\BibitemShut {NoStop}%
\bibitem [{\citenamefont {Lukin}\ and\ \citenamefont {Sotnikov}(2024)}]{3D_Charkiv}%
  \BibitemOpen
  \bibfield  {author} {\bibinfo {author} {\bibfnamefont {I.~V.}\ \bibnamefont {Lukin}}\ and\ \bibinfo {author} {\bibfnamefont {A.~G.}\ \bibnamefont {Sotnikov}},\ }\href {https://doi.org/10.1103/PhysRevB.110.064422} {\bibfield  {journal} {\bibinfo  {journal} {Phys. Rev. B}\ }\textbf {\bibinfo {volume} {110}},\ \bibinfo {pages} {064422} (\bibinfo {year} {2024})}\BibitemShut {NoStop}%
\bibitem [{\citenamefont {Vidal}(2007)}]{Vidal_MERA_07}%
  \BibitemOpen
  \bibfield  {author} {\bibinfo {author} {\bibfnamefont {G.}~\bibnamefont {Vidal}},\ }\href {https://link.aps.org/doi/10.1103/PhysRevLett.99.220405} {\bibfield  {journal} {\bibinfo  {journal} {Phys. Rev. Lett.}\ }\textbf {\bibinfo {volume} {99}},\ \bibinfo {pages} {220405} (\bibinfo {year} {2007})}\BibitemShut {NoStop}%
\bibitem [{\citenamefont {Vidal}(2008)}]{Vidal_MERA_08}%
  \BibitemOpen
  \bibfield  {author} {\bibinfo {author} {\bibfnamefont {G.}~\bibnamefont {Vidal}},\ }\href {https://link.aps.org/doi/10.1103/PhysRevLett.101.110501} {\bibfield  {journal} {\bibinfo  {journal} {Phys. Rev. Lett.}\ }\textbf {\bibinfo {volume} {101}},\ \bibinfo {pages} {110501} (\bibinfo {year} {2008})}\BibitemShut {NoStop}%
\bibitem [{\citenamefont {Evenbly}\ and\ \citenamefont {Vidal}(2014{\natexlab{a}})}]{Evenbly_branchMERA_14}%
  \BibitemOpen
  \bibfield  {author} {\bibinfo {author} {\bibfnamefont {G.}~\bibnamefont {Evenbly}}\ and\ \bibinfo {author} {\bibfnamefont {G.}~\bibnamefont {Vidal}},\ }\href {https://link.aps.org/doi/10.1103/PhysRevLett.112.220502} {\bibfield  {journal} {\bibinfo  {journal} {Phys. Rev. Lett.}\ }\textbf {\bibinfo {volume} {112}},\ \bibinfo {pages} {220502} (\bibinfo {year} {2014}{\natexlab{a}})}\BibitemShut {NoStop}%
\bibitem [{\citenamefont {Evenbly}\ and\ \citenamefont {Vidal}(2014{\natexlab{b}})}]{Evenbly_branchMERAarea_14}%
  \BibitemOpen
  \bibfield  {author} {\bibinfo {author} {\bibfnamefont {G.}~\bibnamefont {Evenbly}}\ and\ \bibinfo {author} {\bibfnamefont {G.}~\bibnamefont {Vidal}},\ }\href {https://link.aps.org/doi/10.1103/PhysRevB.89.235113} {\bibfield  {journal} {\bibinfo  {journal} {Phys. Rev. B}\ }\textbf {\bibinfo {volume} {89}},\ \bibinfo {pages} {235113} (\bibinfo {year} {2014}{\natexlab{b}})}\BibitemShut {NoStop}%
\bibitem [{\citenamefont {Murg}\ \emph {et~al.}(2007)\citenamefont {Murg}, \citenamefont {Verstraete},\ and\ \citenamefont {Cirac}}]{Murg_finitePEPS_07}%
  \BibitemOpen
  \bibfield  {author} {\bibinfo {author} {\bibfnamefont {V.}~\bibnamefont {Murg}}, \bibinfo {author} {\bibfnamefont {F.}~\bibnamefont {Verstraete}},\ and\ \bibinfo {author} {\bibfnamefont {J.~I.}\ \bibnamefont {Cirac}},\ }\href {https://link.aps.org/doi/10.1103/PhysRevA.75.033605} {\bibfield  {journal} {\bibinfo  {journal} {Phys. Rev. A}\ }\textbf {\bibinfo {volume} {75}},\ \bibinfo {pages} {033605} (\bibinfo {year} {2007})}\BibitemShut {NoStop}%
\bibitem [{\citenamefont {Jordan}\ \emph {et~al.}(2008)\citenamefont {Jordan}, \citenamefont {Or\'us}, \citenamefont {Vidal}, \citenamefont {Verstraete},\ and\ \citenamefont {Cirac}}]{Cirac_iPEPS_08}%
  \BibitemOpen
  \bibfield  {author} {\bibinfo {author} {\bibfnamefont {J.}~\bibnamefont {Jordan}}, \bibinfo {author} {\bibfnamefont {R.}~\bibnamefont {Or\'us}}, \bibinfo {author} {\bibfnamefont {G.}~\bibnamefont {Vidal}}, \bibinfo {author} {\bibfnamefont {F.}~\bibnamefont {Verstraete}},\ and\ \bibinfo {author} {\bibfnamefont {J.~I.}\ \bibnamefont {Cirac}},\ }\href {https://link.aps.org/doi/10.1103/PhysRevLett.101.250602} {\bibfield  {journal} {\bibinfo  {journal} {Phys. Rev. Lett.}\ }\textbf {\bibinfo {volume} {101}},\ \bibinfo {pages} {250602} (\bibinfo {year} {2008})}\BibitemShut {NoStop}%
\bibitem [{\citenamefont {Jiang}\ \emph {et~al.}(2008)\citenamefont {Jiang}, \citenamefont {Weng},\ and\ \citenamefont {Xiang}}]{Xiang_SU_08}%
  \BibitemOpen
  \bibfield  {author} {\bibinfo {author} {\bibfnamefont {H.~C.}\ \bibnamefont {Jiang}}, \bibinfo {author} {\bibfnamefont {Z.~Y.}\ \bibnamefont {Weng}},\ and\ \bibinfo {author} {\bibfnamefont {T.}~\bibnamefont {Xiang}},\ }\href {https://link.aps.org/doi/10.1103/PhysRevLett.101.090603} {\bibfield  {journal} {\bibinfo  {journal} {Phys. Rev. Lett.}\ }\textbf {\bibinfo {volume} {101}},\ \bibinfo {pages} {090603} (\bibinfo {year} {2008})}\BibitemShut {NoStop}%
\bibitem [{\citenamefont {Or\'us}\ and\ \citenamefont {Vidal}(2009)}]{Orus_CTM_09}%
  \BibitemOpen
  \bibfield  {author} {\bibinfo {author} {\bibfnamefont {R.}~\bibnamefont {Or\'us}}\ and\ \bibinfo {author} {\bibfnamefont {G.}~\bibnamefont {Vidal}},\ }\href {https://link.aps.org/doi/10.1103/PhysRevB.80.094403} {\bibfield  {journal} {\bibinfo  {journal} {Phys. Rev. B}\ }\textbf {\bibinfo {volume} {80}},\ \bibinfo {pages} {094403} (\bibinfo {year} {2009})}\BibitemShut {NoStop}%
\bibitem [{\citenamefont {Phien}\ \emph {et~al.}(2015)\citenamefont {Phien}, \citenamefont {Bengua}, \citenamefont {Tuan}, \citenamefont {Corboz},\ and\ \citenamefont {Or\'us}}]{fu}%
  \BibitemOpen
  \bibfield  {author} {\bibinfo {author} {\bibfnamefont {H.~N.}\ \bibnamefont {Phien}}, \bibinfo {author} {\bibfnamefont {J.~A.}\ \bibnamefont {Bengua}}, \bibinfo {author} {\bibfnamefont {H.~D.}\ \bibnamefont {Tuan}}, \bibinfo {author} {\bibfnamefont {P.}~\bibnamefont {Corboz}},\ and\ \bibinfo {author} {\bibfnamefont {R.}~\bibnamefont {Or\'us}},\ }\href {https://link.aps.org/doi/10.1103/PhysRevB.92.035142} {\bibfield  {journal} {\bibinfo  {journal} {Phys. Rev. B}\ }\textbf {\bibinfo {volume} {92}},\ \bibinfo {pages} {035142} (\bibinfo {year} {2015})}\BibitemShut {NoStop}%
\bibitem [{\citenamefont {Lubasch}\ \emph {et~al.}(2014)\citenamefont {Lubasch}, \citenamefont {Cirac},\ and\ \citenamefont {Ba\~nuls}}]{Lubasch_conditioning}%
  \BibitemOpen
  \bibfield  {author} {\bibinfo {author} {\bibfnamefont {M.}~\bibnamefont {Lubasch}}, \bibinfo {author} {\bibfnamefont {J.~I.}\ \bibnamefont {Cirac}},\ and\ \bibinfo {author} {\bibfnamefont {M.-C.}\ \bibnamefont {Ba\~nuls}},\ }\href {https://doi.org/10.1103/PhysRevB.90.064425} {\bibfield  {journal} {\bibinfo  {journal} {Phys. Rev. B}\ }\textbf {\bibinfo {volume} {90}},\ \bibinfo {pages} {064425} (\bibinfo {year} {2014})}\BibitemShut {NoStop}%
\bibitem [{\citenamefont {Corboz}(2016{\natexlab{a}})}]{Corboz_varopt_16}%
  \BibitemOpen
  \bibfield  {author} {\bibinfo {author} {\bibfnamefont {P.}~\bibnamefont {Corboz}},\ }\href {https://link.aps.org/doi/10.1103/PhysRevB.94.035133} {\bibfield  {journal} {\bibinfo  {journal} {Phys. Rev. B}\ }\textbf {\bibinfo {volume} {94}},\ \bibinfo {pages} {035133} (\bibinfo {year} {2016}{\natexlab{a}})}\BibitemShut {NoStop}%
\bibitem [{\citenamefont {Vanderstraeten}\ \emph {et~al.}(2016)\citenamefont {Vanderstraeten}, \citenamefont {Haegeman}, \citenamefont {Corboz},\ and\ \citenamefont {Verstraete}}]{Vanderstraeten_varopt_16}%
  \BibitemOpen
  \bibfield  {author} {\bibinfo {author} {\bibfnamefont {L.}~\bibnamefont {Vanderstraeten}}, \bibinfo {author} {\bibfnamefont {J.}~\bibnamefont {Haegeman}}, \bibinfo {author} {\bibfnamefont {P.}~\bibnamefont {Corboz}},\ and\ \bibinfo {author} {\bibfnamefont {F.}~\bibnamefont {Verstraete}},\ }\href {https://link.aps.org/doi/10.1103/PhysRevB.94.155123} {\bibfield  {journal} {\bibinfo  {journal} {Phys. Rev. B}\ }\textbf {\bibinfo {volume} {94}},\ \bibinfo {pages} {155123} (\bibinfo {year} {2016})}\BibitemShut {NoStop}%
\bibitem [{\citenamefont {Fishman}\ \emph {et~al.}(2018)\citenamefont {Fishman}, \citenamefont {Vanderstraeten}, \citenamefont {Zauner-Stauber}, \citenamefont {Haegeman},\ and\ \citenamefont {Verstraete}}]{Fishman_FPCTM_17}%
  \BibitemOpen
  \bibfield  {author} {\bibinfo {author} {\bibfnamefont {M.~T.}\ \bibnamefont {Fishman}}, \bibinfo {author} {\bibfnamefont {L.}~\bibnamefont {Vanderstraeten}}, \bibinfo {author} {\bibfnamefont {V.}~\bibnamefont {Zauner-Stauber}}, \bibinfo {author} {\bibfnamefont {J.}~\bibnamefont {Haegeman}},\ and\ \bibinfo {author} {\bibfnamefont {F.}~\bibnamefont {Verstraete}},\ }\href {https://link.aps.org/doi/10.1103/PhysRevB.98.235148} {\bibfield  {journal} {\bibinfo  {journal} {Phys. Rev. B}\ }\textbf {\bibinfo {volume} {98}},\ \bibinfo {pages} {235148} (\bibinfo {year} {2018})}\BibitemShut {NoStop}%
\bibitem [{\citenamefont {Xie}\ \emph {et~al.}(2017)\citenamefont {Xie}, \citenamefont {Liao}, \citenamefont {Huang}, \citenamefont {Xie}, \citenamefont {Chen}, \citenamefont {Liu},\ and\ \citenamefont {Xiang}}]{Xie_PEPScontr_17}%
  \BibitemOpen
  \bibfield  {author} {\bibinfo {author} {\bibfnamefont {Z.~Y.}\ \bibnamefont {Xie}}, \bibinfo {author} {\bibfnamefont {H.~J.}\ \bibnamefont {Liao}}, \bibinfo {author} {\bibfnamefont {R.~Z.}\ \bibnamefont {Huang}}, \bibinfo {author} {\bibfnamefont {H.~D.}\ \bibnamefont {Xie}}, \bibinfo {author} {\bibfnamefont {J.}~\bibnamefont {Chen}}, \bibinfo {author} {\bibfnamefont {Z.~Y.}\ \bibnamefont {Liu}},\ and\ \bibinfo {author} {\bibfnamefont {T.}~\bibnamefont {Xiang}},\ }\href {https://link.aps.org/doi/10.1103/PhysRevB.96.045128} {\bibfield  {journal} {\bibinfo  {journal} {Phys. Rev. B}\ }\textbf {\bibinfo {volume} {96}},\ \bibinfo {pages} {045128} (\bibinfo {year} {2017})}\BibitemShut {NoStop}%
\bibitem [{\citenamefont {Corboz}(2016{\natexlab{b}})}]{Corboz_Eextrap_16}%
  \BibitemOpen
  \bibfield  {author} {\bibinfo {author} {\bibfnamefont {P.}~\bibnamefont {Corboz}},\ }\href {https://link.aps.org/doi/10.1103/PhysRevB.93.045116} {\bibfield  {journal} {\bibinfo  {journal} {Phys. Rev. B}\ }\textbf {\bibinfo {volume} {93}},\ \bibinfo {pages} {045116} (\bibinfo {year} {2016}{\natexlab{b}})}\BibitemShut {NoStop}%
\bibitem [{\citenamefont {Corboz}\ \emph {et~al.}(2018)\citenamefont {Corboz}, \citenamefont {Czarnik}, \citenamefont {Kapteijns},\ and\ \citenamefont {Tagliacozzo}}]{Corboz_FCLS_18}%
  \BibitemOpen
  \bibfield  {author} {\bibinfo {author} {\bibfnamefont {P.}~\bibnamefont {Corboz}}, \bibinfo {author} {\bibfnamefont {P.}~\bibnamefont {Czarnik}}, \bibinfo {author} {\bibfnamefont {G.}~\bibnamefont {Kapteijns}},\ and\ \bibinfo {author} {\bibfnamefont {L.}~\bibnamefont {Tagliacozzo}},\ }\href {https://link.aps.org/doi/10.1103/PhysRevX.8.031031} {\bibfield  {journal} {\bibinfo  {journal} {Phys. Rev. X}\ }\textbf {\bibinfo {volume} {8}},\ \bibinfo {pages} {031031} (\bibinfo {year} {2018})}\BibitemShut {NoStop}%
\bibitem [{\citenamefont {Rader}\ and\ \citenamefont {L\"auchli}(2018)}]{Rader_FCLS_18}%
  \BibitemOpen
  \bibfield  {author} {\bibinfo {author} {\bibfnamefont {M.}~\bibnamefont {Rader}}\ and\ \bibinfo {author} {\bibfnamefont {A.~M.}\ \bibnamefont {L\"auchli}},\ }\href {https://link.aps.org/doi/10.1103/PhysRevX.8.031030} {\bibfield  {journal} {\bibinfo  {journal} {Phys. Rev. X}\ }\textbf {\bibinfo {volume} {8}},\ \bibinfo {pages} {031030} (\bibinfo {year} {2018})}\BibitemShut {NoStop}%
\bibitem [{\citenamefont {Rams}\ \emph {et~al.}(2018)\citenamefont {Rams}, \citenamefont {Czarnik},\ and\ \citenamefont {Cincio}}]{Rams_xiD_18}%
  \BibitemOpen
  \bibfield  {author} {\bibinfo {author} {\bibfnamefont {M.~M.}\ \bibnamefont {Rams}}, \bibinfo {author} {\bibfnamefont {P.}~\bibnamefont {Czarnik}},\ and\ \bibinfo {author} {\bibfnamefont {L.}~\bibnamefont {Cincio}},\ }\href {https://link.aps.org/doi/10.1103/PhysRevX.8.041033} {\bibfield  {journal} {\bibinfo  {journal} {Phys. Rev. X}\ }\textbf {\bibinfo {volume} {8}},\ \bibinfo {pages} {041033} (\bibinfo {year} {2018})}\BibitemShut {NoStop}%
\bibitem [{\citenamefont {Hasik}\ and\ \citenamefont {Becca}(2019)}]{Hasik}%
  \BibitemOpen
  \bibfield  {author} {\bibinfo {author} {\bibfnamefont {J.}~\bibnamefont {Hasik}}\ and\ \bibinfo {author} {\bibfnamefont {F.}~\bibnamefont {Becca}},\ }\href {https://link.aps.org/doi/10.1103/PhysRevB.100.054429} {\bibfield  {journal} {\bibinfo  {journal} {Phys. Rev. B}\ }\textbf {\bibinfo {volume} {100}},\ \bibinfo {pages} {054429} (\bibinfo {year} {2019})}\BibitemShut {NoStop}%
\bibitem [{\citenamefont {Haghshenas}\ \emph {et~al.}(2019)\citenamefont {Haghshenas}, \citenamefont {O'Rourke},\ and\ \citenamefont {Chan}}]{canonical_PEPS}%
  \BibitemOpen
  \bibfield  {author} {\bibinfo {author} {\bibfnamefont {R.}~\bibnamefont {Haghshenas}}, \bibinfo {author} {\bibfnamefont {M.~J.}\ \bibnamefont {O'Rourke}},\ and\ \bibinfo {author} {\bibfnamefont {G.~K.-L.}\ \bibnamefont {Chan}},\ }\href {https://doi.org/10.1103/PhysRevB.100.054404} {\bibfield  {journal} {\bibinfo  {journal} {Phys. Rev. B}\ }\textbf {\bibinfo {volume} {100}},\ \bibinfo {pages} {054404} (\bibinfo {year} {2019})}\BibitemShut {NoStop}%
\bibitem [{\citenamefont {Sokolov}\ \emph {et~al.}(2025)\citenamefont {Sokolov}, \citenamefont {Zhang},\ and\ \citenamefont {Dziarmaga}}]{loops1}%
  \BibitemOpen
  \bibfield  {author} {\bibinfo {author} {\bibfnamefont {I.}~\bibnamefont {Sokolov}}, \bibinfo {author} {\bibfnamefont {Y.}~\bibnamefont {Zhang}},\ and\ \bibinfo {author} {\bibfnamefont {J.}~\bibnamefont {Dziarmaga}},\ }\href {https://doi.org/10.1103/4lgp-ld2s} {\bibfield  {journal} {\bibinfo  {journal} {Phys. Rev. E}\ }\textbf {\bibinfo {volume} {112}},\ \bibinfo {pages} {055307} (\bibinfo {year} {2025})}\BibitemShut {NoStop}%
\bibitem [{\citenamefont {Czarnik}\ \emph {et~al.}(2019)\citenamefont {Czarnik}, \citenamefont {Dziarmaga},\ and\ \citenamefont {Corboz}}]{CzarnikDziarmagaCorboz}%
  \BibitemOpen
  \bibfield  {author} {\bibinfo {author} {\bibfnamefont {P.}~\bibnamefont {Czarnik}}, \bibinfo {author} {\bibfnamefont {J.}~\bibnamefont {Dziarmaga}},\ and\ \bibinfo {author} {\bibfnamefont {P.}~\bibnamefont {Corboz}},\ }\href {https://doi.org/10.1103/PhysRevB.99.035115} {\bibfield  {journal} {\bibinfo  {journal} {Phys. Rev. B}\ }\textbf {\bibinfo {volume} {99}},\ \bibinfo {pages} {035115} (\bibinfo {year} {2019})}\BibitemShut {NoStop}%
\bibitem [{\citenamefont {Dziarmaga}(2021)}]{ntu}%
  \BibitemOpen
  \bibfield  {author} {\bibinfo {author} {\bibfnamefont {J.}~\bibnamefont {Dziarmaga}},\ }\href {https://doi.org/10.1103/PhysRevB.104.094411} {\bibfield  {journal} {\bibinfo  {journal} {Phys. Rev. B}\ }\textbf {\bibinfo {volume} {104}},\ \bibinfo {pages} {094411} (\bibinfo {year} {2021})}\BibitemShut {NoStop}%
\bibitem [{\citenamefont {Sinha}\ \emph {et~al.}(2022)\citenamefont {Sinha}, \citenamefont {Rams}, \citenamefont {Czarnik},\ and\ \citenamefont {Dziarmaga}}]{Hubbard_Sinha}%
  \BibitemOpen
  \bibfield  {author} {\bibinfo {author} {\bibfnamefont {A.}~\bibnamefont {Sinha}}, \bibinfo {author} {\bibfnamefont {M.~M.}\ \bibnamefont {Rams}}, \bibinfo {author} {\bibfnamefont {P.}~\bibnamefont {Czarnik}},\ and\ \bibinfo {author} {\bibfnamefont {J.}~\bibnamefont {Dziarmaga}},\ }\href {https://doi.org/10.1103/PhysRevB.106.195105} {\bibfield  {journal} {\bibinfo  {journal} {Phys. Rev. B}\ }\textbf {\bibinfo {volume} {106}},\ \bibinfo {pages} {195105} (\bibinfo {year} {2022})}\BibitemShut {NoStop}%
\bibitem [{\citenamefont {Sinha}\ and\ \citenamefont {Wietek}(2024)}]{Sinha_Wietek_Hubbard}%
  \BibitemOpen
  \bibfield  {author} {\bibinfo {author} {\bibfnamefont {A.}~\bibnamefont {Sinha}}\ and\ \bibinfo {author} {\bibfnamefont {A.}~\bibnamefont {Wietek}},\ }\href {https://arxiv.org/abs/2411.15158} {\bibinfo {title} {Forestalled phase separation as the precursor to stripe order}} (\bibinfo {year} {2024}),\ \Eprint {https://arxiv.org/abs/2411.15158} {arXiv:2411.15158 [cond-mat.str-el]} \BibitemShut {NoStop}%
\bibitem [{\citenamefont {Zhang}\ \emph {et~al.}(2026)\citenamefont {Zhang}, \citenamefont {Sinha}, \citenamefont {Rams},\ and\ \citenamefont {Dziarmaga}}]{tJ_Zhang}%
  \BibitemOpen
  \bibfield  {author} {\bibinfo {author} {\bibfnamefont {Y.}~\bibnamefont {Zhang}}, \bibinfo {author} {\bibfnamefont {A.}~\bibnamefont {Sinha}}, \bibinfo {author} {\bibfnamefont {M.~M.}\ \bibnamefont {Rams}},\ and\ \bibinfo {author} {\bibfnamefont {J.}~\bibnamefont {Dziarmaga}},\ }\href {https://doi.org/10.1103/6pcg-qq4p} {\bibfield  {journal} {\bibinfo  {journal} {Phys. Rev. B}\ }\textbf {\bibinfo {volume} {113}},\ \bibinfo {pages} {085113} (\bibinfo {year} {2026})}\BibitemShut {NoStop}%
\bibitem [{\citenamefont {Ponsioen}\ and\ \citenamefont {Corboz}(2020)}]{gapCorboz}%
  \BibitemOpen
  \bibfield  {author} {\bibinfo {author} {\bibfnamefont {B.}~\bibnamefont {Ponsioen}}\ and\ \bibinfo {author} {\bibfnamefont {P.}~\bibnamefont {Corboz}},\ }\href {https://doi.org/10.1103/PhysRevB.101.195109} {\bibfield  {journal} {\bibinfo  {journal} {Phys. Rev. B}\ }\textbf {\bibinfo {volume} {101}},\ \bibinfo {pages} {195109} (\bibinfo {year} {2020})}\BibitemShut {NoStop}%
\end{thebibliography}%



\end{document}